# Tumor-Centered Patching for Enhanced Medical Image Segmentation

Mutyyba Asghar[1], Ahmad Raza Shahid[1], Akhtar Jamil[1], Kiran Aftab[2], Syed Ather Enam[2]

[1] National University of Computer and Emerging Sciences
[2] The Aga Khan University

**Abstract**
The realm of medical image diagnosis has advanced significantly with the integration of computer-aided diagnosis and surgical systems. However, challenges persist, particularly in achieving precise image segmentation. While deep learning techniques show potential, obstacles like limited resources, slow convergence, and class imbalance impede their effectiveness. Traditional patch-based methods, though common, struggle to capture intricate tumor boundaries and often lead to redundant samples, compromising computational efficiency and feature quality. To tackle these issues, this research introduces an innovative approach centered on the tumor itself for patch-based image analysis. This novel tumor-centered patching method aims to address the class imbalance and boundary deficiencies, enabling focused and accurate tumor segmentation. By aligning patches with the tumor's anatomical context, this technique enhances feature extraction accuracy and reduces computational load. Experimental results demonstrate improved class imbalance, with segmentation scores of 0.78, 0.76, and 0.71 for whole, core, and enhancing tumors, respectively using a lightweight simple U-Net. This approach shows potential for enhancing medical image segmentation and improving computer-aided diagnosis systems.

## 1. Introduction

In the context of adopting the "whole-image" strategy, it's important to take into account a couple of limitations. One of the main issues is a class imbalance, which can cause the learning process to ignore infrequent tumor sub-regions. Additionally, treating the entire image as a single entity is similar to processing numerous patches in a batch (Ning et al., 2005), which can lead to slower convergence rates hence increasing the use of computational resources. Previous research has shown that updating weights after each sample can be a more efficient approach than updating weights after accumulating gradients over a batch of samples (LeCun et al., 2002).

In deep learning dealing with large datasets such as Heart, aorta, trachea, and esophagus in CT images (Simpson et al., 2019) lung nodules in CT images (Landman et al., 2015), Liver, spleen, right and left kidney MRI scans (Kavur et al., 2021), kidney and kidney tumors in CT images (Heller et al., 2019), patching can be useful to overcome memory limitations in GPUs or CPUs (Wang et al., 2020) Because deep learning models need a large amount of data (Aggarwal, 2018). This technique involves dividing the large 2D images and 3D volumes into smaller patches which can be processed in parallel, making it more manageable.

Patching can make a difference in terms of accuracy and efficiency, according to (Long et al., 2015) this can be achieved by randomly sampling patches from the original image. Patching helps better represent the diversity of the original image in the training data, allowing the model to learn more robust features and improve its performance on unseen data. Another great benefit of patching is that it can help balance the distribution of classes in a dataset. Data sampling is particularly useful when certain classes have fewer examples than others, leading to class imbalance. By sampling patches from the image, it is possible to ensure that each patch contains a balanced number of examples from each class.

When it comes to patch extraction, various methods can be used. However, the most commonly used techniques include cropping. This method involves selecting a dominant patch from each image with a specific size for height and width. Random cropping is also often used as it helps to reduce the input size. For example, the size of an image can be reduced from (240, 240) to (128, 128). However, it is important to note that the label-preserving transformation may not be addressed depending on the selected reduction threshold for cropping (Alzubaidi et al., 2021).

In previous studies, researchers found that many people also used non-uniform windows (Ciresan et al., 2012). The reason for this is that when dealing with larger window sizes of uniform patching, it becomes more difficult to accurately capture all the necessary details. Non-uniform sampling allows for more flexibility in choosing the areas that need to be analyzed in greater detail, which can ultimately lead to extracting features at different spatial resolutions hence resulting in more accurate results. The central part of the window is responsible for capturing the finer details of an image, while the periphery of the window captures a more general and coarse representation (Ciresan et al., 2012). However, there are also challenges associated with using non-uniform windows, such as the potential for missing important information in areas that are not selected for analysis. The larger window sizes generally improve performance in various applications. However, using large window sizes also results in more complex networks, which require more computational resources and training data for effective generalization.

Primary and secondary brain tumors represent two distinct types of brain malignancies. The former originates from brain cells, while the latter results from the spread of tumors from other organs to the brain. Gliomas, a specific form of primary brain tumor, can be further categorized into low-grade (LGG) or high-grade (HGG) types. High-grade gliomas are notably aggressive and malignant tumors, often requiring surgical intervention and radiotherapy, and they are associated with an unfavorable prognosis for survival. Among primary brain malignancies, gliomas are the most prevalent, displaying a range of aggressiveness, prognostic outcomes, and diverse histological sub-regions. These sub-regions encompass peritumoral edema, a necrotic core, an enhancing tumor core, and a non-enhancing tumor core. The inherent variability of gliomas is also evident in their radiographic attributes, as multiparametric MRI (mpMRI) scans reveal distinct intensity profiles, indicative of underlying differences in tumor biology.

In the field of diagnosing and treating brain tumors, medical imaging plays a crucial role. Magnetic resonance imaging (MRI) is widely utilized among various imaging modalities for brain tumor examination. Detecting brain tumors in MRI images is an arduous and time-consuming task for radiologists. It demands a high level of expertise and attention to detail. To address this challenge, automatic segmentation techniques have been developed to alleviate the limitations of manual extraction.

The implementation of automatic segmentation algorithms in brain tumor diagnosis not only saves time but also ensures consistent and reliable results. These algorithms leverage sophisticated image processing techniques and machine learning algorithms to segment brain tumors from MRI images. Various approaches have been proposed, such as traditional threshold-based techniques (Akram & Usman, 2011; Husham et al., 2020), clustering-based algorithms (Nasor & Obaid, 2020), and deep learning-based methods (Baid et al., 2021; Sharif et al., 2020).

The purpose of the Brain Tumor Segmentation Challenge (BraTS) 2021 was to facilitate comparison and assessment of diverse automated segmentation algorithms. These algorithms were tested using pre-operative MRI scans obtained from multiple institutions, with the aim of segmenting the inherently diverse sub-regions within brain tumors (Baid et al., 2021). Specifically, the challenge dataset encompasses pre-operative multimodal MRI scans of both glioblastoma (GBM/HGG) and low-grade glioma (LGG), incorporating various modalities such as native (T1), post-contrast T1-weighted (T1Gd), T2-weighted (T2), and Fluid Attenuated Inversion Recovery (FLAIR) volumes.

A 3D UNet is a popular CNN architecture for automatic brain tumor segmentation (Isensee et al., 2018). The multi-scale contextual information of the encoder-decoder sub-networks is effective for the accurate brain tumor segmentation task. Several variations of the encoder-decoder architectures were proposed for MICCAI BraTS 2012-2022 competitions. The potential of several deep architectures (Kamnitsas et al., 2017; Long et al., 2015; Ronneberger et al., 2015) and their ensemble procedures for brain tumor segmentation was discussed by a top-performing method (Kamnitsas et al., 2018) for MICCAI BRATS 2017 competition. Most of the studies use patches for training to reduce the training time and computational complexity. Random crop (Feng et al., 2022), random biased crop (Feng et al., 2022; Futrega et al., 2022), overlapping (Akil et al., 2020), and fixed patches (Beers et al., 2017) are mostly used for model training. The randomly cropped samples may not contain the whole tumor and miss the boundaries of the tumor. In contrast, in a random biased crop, there is the need for ground truth to extract the patch containing the foreground pixels exactly. In overlapping patches, as the Brain MRI data is highly unbalanced, the large number of patches does not contain the tumor, which causes a drastic increase in training time and the computational burden. The performance of the patch-based method may be influenced by the size and location of the patch; cropping each case into a volume of $176 \times 176 \times 155$, and removing the border area, may still result in more background pixels (Huang et al., 2021). So there is a need for an effective patching method that contains more foreground pixels containing tumor by preserving the boundaries and is data independent, and cause a reduction in the computational burden for more effective training. In many of the studies, tumor localization is performed prior to patching (Wang et al., 2022), which is accomplished by training a model that adds to the computational burden.

This paper centers on extracting patches from regions of interest, particularly emphasizing utilizing topological data analysis (TDA) and Connected Component Analysis (CCA) based approaches. Implementing these methods is crucial for optimizing the selection of patch size and placement to cover the entirety of the tumor region. This patching technique ensures accurate capture of the tumor's extent and feature extraction from the region of interest. Additionally, the proposed approach mitigates the influence of data imbalance, thereby improving the robustness and reliability of diagnostic outcomes. In summary, the utilization of TDA-based approaches substantially improves the accuracy and dependability of MRI-based tumor diagnosis and treatment recommendations. The following are the contributions of this study.

- The proposed methodology introduces a novel approach that leverages a label-preserving transformation for accurate patch extraction by identifying the entire tumor as the region of interest. It utilizes topological data analysis (TDA) to extract clusters of persistence images for each connected component, along with their respective locations, for patch extraction.
- Acquisition of patches having tumors near to the center of the patch, hence preserving boundaries and overall shape of the tumor.
- A targeted strategy that emphasizes tumor pixels while balancing the presence of healthy brain pixels was used to tackle the class imbalance between tumorous and healthy cells. The approach was evaluated by calculating the background and tumorous pixels ratio.

## 2. Literature Review

The literature review is conducted within the framework of the patching technique tailored for brain tumor segmentation. Diverse methods and models have been put forth, each employing different patching techniques. These techniques, predominantly employed in brain tumor segmentation, encompass fixed-sized patches, random patches, random biased patches, and overlapping patches.

### 2.1. Fixed Sized Patches

One common approach in brain tumor segmentation involves dividing the MRI image into small fixed-sized patches. While this method simplifies the segmentation task, it has its limitations. For instance, it may not capture the global context of the tumor, resulting in inaccurate segmentations for tumors with irregular shapes or locations. Additionally, fixed-sized patches can sometimes lead to disjointed tumor regions, making it challenging to merge the segments and obtain complete tumor segmentation. The accuracy of this approach is highly dependent on the patch size and location used, and larger tumors may require more complex segmentation techniques. Smaller patches may also necessitate additional post-processing steps to achieve accurate tumor segmentation. Similarly in this research(Peiris et al., 2022), the author has taken a patch of 128x128x128 from a brain MRI for reciprocal adversarial learning that performs well on all tumor labels. Dice Similarity Scores of 81.38%, 90.77%, and 85.39%; Hausdorff Distances (95%) of 21.83 mm, 5.37 mm, and 8.56 mm were obtained for the enhancing tumor, whole tumor, and tumor core, respectively.

A new approach is presented in this research for brain MRI segmentation called the patch-wise U-net architecture, which aims to overcome the limitations of the conventional U-net model. One of the strengths of this approach is that it better retains local spatial information and is capable of handling multi-class segmentation. The experimental results show that the proposed model outperforms previously proposed methods, as demonstrated by the high Dice similarity coefficient and Jacquard index scores on the OASIS and IBSR datasets. However, the method has a weakness, which is the increased computational complexity during training, and also the method adopted to extract tumor patches is not efficient. The whole image is converted into four equal patches. Thus some patches may have more class imbalance problems (Lee et al., 2020).

## 2.2. Random Patches
An alternative technique involves using random patches for brain tumor segmentation. One study proposes a BiTr-Unet model, combining convolutional neural networks (CNNs) and vision transformers for leveraging the strengths of both architectures and trained on patches of data. Furthermore, the model performs well on the BraTS2021 testing dataset, with Dice scores of 0.9257, 0.9350, and 0.8874, for the whole tumor, tumor core, and enhancing tumor, respectively. (Jia & Shu, 2022)
Similarly, another research, (Ghaffari et al., 2021) introduces a deep learning-based method for segmenting brain tumors into sub-regions. They utilize a 3D CNN combined with patch extraction in their approach to tackle class imbalance and accurately localize tumor. The model demonstrates impressive Dice Scores for different tumor regions, showcasing the efficacy of their approach. Another research (Kazerooni et al., 2023)has employed the GANDLF (Pati et al., 2021) model that also has trained the different models with different patch sizes for pediatric brain tumor segmentation.

Additionally, a study employed a combination of 2D and 3D networks for tumor segmentation due to GPU memory limitations. The use of smaller patches allowed them to leverage a standard UNet-3D model, achieving good results with potential implications for survival prediction.
Furthermore, the scale-attention network was used to achieve high accuracy in tumor segmentation. The model incorporates a dynamic scale attention mechanism and was trained on patches of a large dataset, obtaining notable Dice Similarity Coefficient scores for various tumor regions (Yuan, 2022).
(Yan et al., 2022) Proposed a modified U-net model to enhance glioblastoma segmentation. The model reduces computational resource requirements by utilizing 2D convolutions and 2D patched data. This technique allows for more efficient and affordable region segmentation, as it narrows down the search space for the model. However, it is important to note that the model's use of 2D data lacks the comprehensive 3D aspect of tumor appearance. Tumor characteristics can vary in 3D, and the absence of this information may limit the accuracy of the model's predictions.

## 2.3. Random Biased Patches
A study presented an optimized U-Net architecture for brain tumor segmentation in the BraTS21 challenge. The authors conducted an extensive ablation study and utilized random biased crop augmentation to achieve improved segmentation results (Futrega et al., 2022).
In their study (Ahmad et al., 2022) introduced an innovative approach to improve image segmentation accuracy. Their method incorporates three key strategies: densely connected blocks for capturing multi-scale features, residual-inception blocks for extracting local and global information, and deep supervision for effective gradient propagation during training. Evaluating the MS UNet model using 3D random patches from the BraTS 2021 validation dataset, the authors achieved remarkable accuracy in tumor segmentation. The results indicated Dice scores of 91.938% for the entire tumor, 86.268% for the tumor core, and 82.409% for the enhancing tumor. This study demonstrates the efficacy of their multi-scale strategy in enhancing image segmentation accuracy, particularly in the context of tumor segmentation.

## 2.4. Overlapping Patches

An overlapping patch-based technique involves dividing the MRI image into small overlapping patches to capture more contextual information.While this approach has advantages, such as improved context, it also comes with certain limitations. Increased computational complexity is one of the drawbacks, as it involves processing multiple overlapping patches for each image region. This may lead to overfitting and more false positives, necessitating additional post-processing for complete and accurate tumor segmentation. In the research conducted by (Ullah et al., 2023), brain tumor segmentation is performed using a patch-based convolutional neural network (CNN) approach. The study incorporates early-stage classification to differentiate between tumor and non-tumor slices. Various images affected by noise and artifacts are discarded to ensure data quality. The segmentation task is carried out on the BRATS dataset from 2012 to 2018. Subsequently, overlapping patches of size 64x64x64 are extracted to segment the tumors effectively. The proposed model achieves a dice score of 0.91 for brain tumor segmentation. This approach presents a promising methodology for accurately segmenting brain tumors, leveraging patch-based CNNs and early-stage classification techniques.

In another research (Huang et al., 2021), a multi-scale feature fusing architecture was proposed, combining two main networks: the Feature Extraction Network (FEN) and the Multi-scale Feature Fusing Network (MSFFN). This approach demonstrated the ability to capture more detailed and semantic information, leading to more precise segmentation results. However, it increases computational complexity during training and may be sensitive to patch size and location. , as well as the overlap between adjacent patches. Finally, the increased GPU memory cost during training may lead to slower GPU inference speeds than other approaches, such as U-net.

Similarly in this research (Islam et al., 2022), the author has synthetically generated the CT scans from the 3D MRI to avoid the co-registration problem. All 4 modalities of the newly generated CT scan are stacked together to process through the UNET. In this paper, the author has also performed, input-level fusion, decision-level fusion, and layer-level fusion. The dataset employed in this study was Brats 2021. Different loss functions have been used in this study to optimize the loss calculation such as Tversky and generalized dice loss function. The mean IOU achieved in this study was about 0.84.

Another study utilized an occipital temporal pathway in a CNN to address the class imbalance and selectively attend to overlapping patches for segmenting small tumors with fuzzy borders. The proposed method achieved promising results, demonstrating the potential of deep learning for accurately segmenting challenging tumors. (Akil et al., 2020)

In conclusion, various patch-based techniques have been explored for brain tumor segmentation. Each approach comes with its strengths and limitations, impacting accuracy, efficiency, and computational complexity. Our proposed model addresses these challenges and aims to achieve improved segmentation results while minimizing drawbacks observed in existing techniques.

 our research presents a novel patching technique that addresses two critical challenges in tumor segmentation. Firstly, it efficiently forms tumor-centered patches, effectively reducing class imbalance in the dataset. Secondly, it enables the use of the U-net architecture with significantly fewer parameters, optimizing performance without compromising accuracy.

## 3. Proposed Methodology

This paper mainly focuses on extracting patches to reduce computational workload efficiently. The goal is to capture the entire tumor within the patches, allowing the model to concentrate on the most important voxels and pixels. This method helps extract complete tumor features while preserving boundaries, often lost when patches are randomly cropped. The paper outlines the steps involved in extracting 2D and 3D patches. The overall flow of the paper is shown in this figure 1.

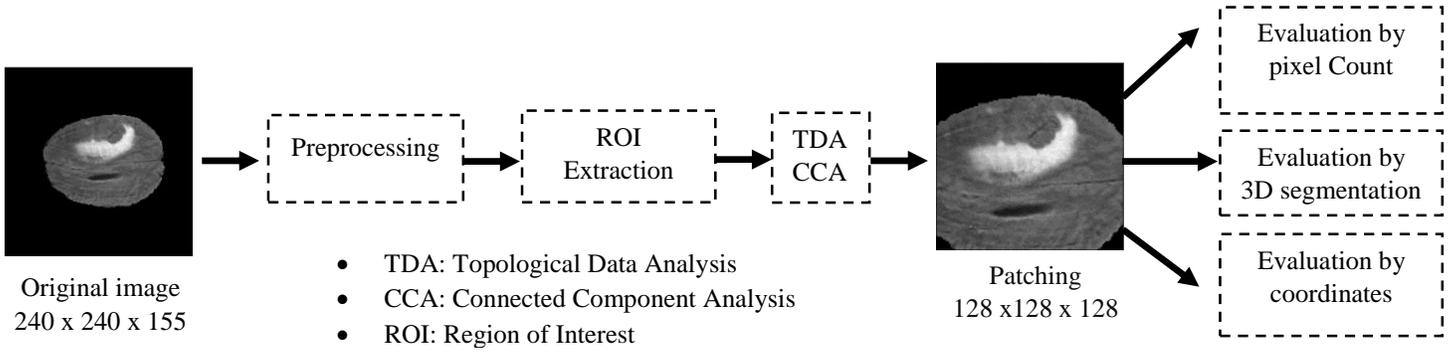

*Figure 1: Overall Work flow for Patching and its evaluation*

### 3.1. Preprocessing

As part of the preprocessing procedures, the 3D volume's mean and standard deviation are adjusted to achieve a mean of zero and a standard deviation of one, as outlined in equation 1. After normalization, the volumes are converted into two-dimensional slices for further processing. Flair volume is considered for patching as it is a modified version of T2 weighted. The lesion appears hyper intense on T2-weighted (T2w) magnetic resonance (MR) images, and their quantification is an important biomarker for the diagnosis and follow-up of the disease (Carass et al., 2017).

$$x_p = \frac{(x-\bar{x})}{\sigma} \qquad (1)$$

Here, $x_p$ denotes the preprocessed 3D volume, while $x$ represents the input 3D volume. Additionally, $\bar{x}$ stands for the mean value of the 3D volume. On the other hand, $\sigma$ corresponds to the standard deviation of the input.

### 3.2. Region of Interest Extraction

In this stage, the data preprocessing is done through a series of sequential steps. Firstly, guassian blur is applied to the images to eliminate small holes as shown in step 2 of Figure 2, and this is achieved by employing a linear spatial filter, as equation 2 depicts. Next, the edges of the image are extracted using a custom cross filter, which applies high-pass spatial filtering to obtain vertical and horizontal edge gradients and their directions using equations 3, 4 and, 5. These extracted edges are added to the original image, enhancing its sharpness as shown in step 3 of figure 2. Subsequently, automatic multilevel thresholding (Yen et al., 1995) is performed on the image histogram to segment the region with the highest intensities as shown in step 4 of figure 2, as the tumor typically appears as the brightest region in the flair modality. The segmented region is then undergo by morphological operations to remove small components as shown in step 5 of figure 2. The region of interest is further subjected to topological data analysis for subsequent processing.

$$g(x,y) = \sum_{s=-a}^{a} \sum_{t=-b}^{b} \omega(s,t) f(x+s, y+t) \qquad (2)$$

Equation 1 shows the linear spatial filtering of an image of size (x,y) with a kernel of size (s,t).

$$gradient\ of\ image = \nabla f = \left[\frac{\partial f}{\partial x}, \frac{\partial f}{\partial y}\right] \qquad (3)$$

$$gradient\ direction = \theta = \tan^{-1}\left[\frac{\partial f}{\partial x}, \frac{\partial f}{\partial y}\right] \qquad (4)$$

$$gradient\ magnitude = \|\nabla f\| = \sqrt{\left(\frac{\partial f}{\partial f}\right)^2 + \left(\frac{\partial f}{\partial y}\right)^2} \qquad (5)$$

This symbol ∇f represents the gradient of the image. It is a vector that contains two components the partial derivative of the image function f concerning the x-coordinate (∂f/∂x) and the partial derivative of the image function f concerning the y-coordinate (∂f/∂y). Whereas $\theta$ represents the direction of the gradient.

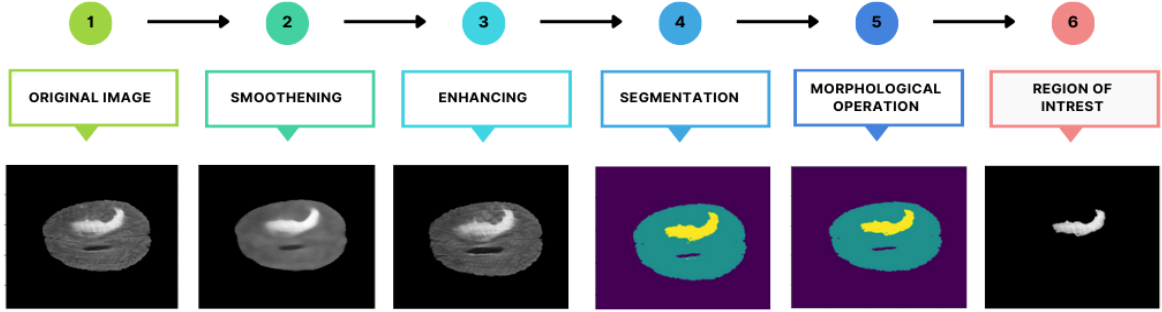

*Figure 2: Pipeline for the region of interest segmentation (1) Original image, (2) Applied Gaussian Blur, (3) Image enhancement, (4) Thresholding, (5)Applied Opening and closing, (6) Extracted ROI*

### 3.3. Topological Data Analysis

Topology examines the structure, distribution, and relationships of data points. In this context, Topological Data Analysis (TDA) assists in extracting shape-related features from the data. While the human eye can typically perceive global patterns in data, TDA can capture both local and global aspects (Demir et al., 2023; Keros et al., 2022; Moon et al., 2020). These features manifest as sets of homology groups, which can be thought of intuitively as connected component, holes or voids in the desired dimension. Persistence homology examines the shape of the extracted region of interest. This technique enables the analysis of persistent topological features in the data, providing insights into the shape characteristics of the region.

### 3.4. 1 Persistent Homology

Persistent homology is a mathematical technique employed in topological data analysis for investigating the shape characteristics of data. It involves creating a filtration of the dataset, a series of nested subsets, using cubical complexes (Moon et al., 2020). These complexes gradually unveil the underlying shape of the data. The cubical complexes are constructed using filtration values, enabling a comprehensive exploration of the data's shape properties (Kaji et al., 2020). Let $C_\varepsilon$ be the cubical complex with filtration $\varepsilon$. In $C_\varepsilon$, the pixels whose assigned values are less than $\varepsilon$ enter the complex. Figure 3 present the sequence of cubical complexes with $\varepsilon \in$ (pixel intensity).

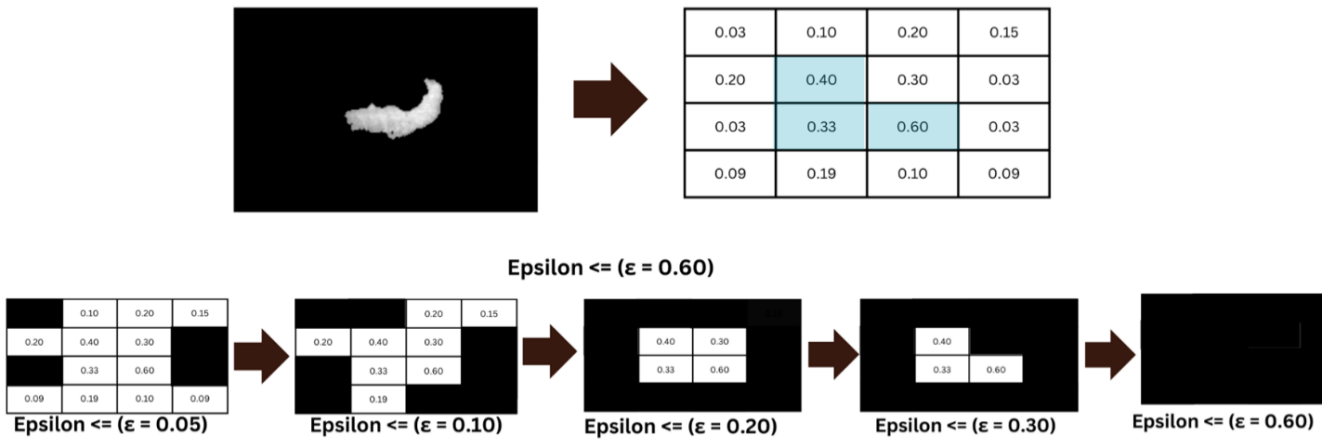

*Figure 3: Nested sub-sequences of Cubical complexes in Persistent Homology*

The persistence diagram provides valuable information regarding the persistence of connected components and holes in the data. It displays the number of cycles, their dimensions, and the birth and death of each homological class. Through analysis of the persistence diagram, we can obtain information regarding the quantity and positioning of connected components, holes, and voids within the area of focus. In the proposed methodology, we specifically focus on utilizing the connected components from the output of the persistence homology for subsequent processing steps.

### 3.5. Persistence Images

Homology is a mathematical concept that captures the topological features of a space. The information it encodes pertains to the connectivity and holes in space globally and locally. It also includes details about the precise location of these features. When using the Cubical Ripser software(Kaji et al., 2020) to compute the homology of an image, we can extract the birth and death coordinates of each homological feature, such as a connected component. These coordinates provide rough localization information about where the connected component exists in the image. Furthermore, we can use this information to create another image for each homology dimension showing the maximum lifetime of the connected component at each pixel location. The lifetime of a cycle is defined as its death time minus its birth time. This image is sparse, with many pixels having a zero value, as not all locations contain homological features. This image can be used with other image analysis techniques, such as deep learning, to analyze the image further and extract more meaningful information. Combining the homological information with other techniques can give us a more comprehensive understanding of the image's structure and features. This image contains the persistence image of each connected component in the obtained image. This feature representation is also called the persistence surface function. A persistence surface function as shown in equation 6 is a stable representation of a persistence diagram; persistence surface functions are robust to small perturbations of points in persistence diagrams (Adams et al., 2017).

$$\rho p(x, y) = \sum_{(b,d) \in P} g_{(b,d)}(x, y) . \omega(b, d) \qquad (6)$$

Where, x and y are the (x,y)-coordinates of the persistence function, $g_{(b,d)}$ is a smoothening function for $(b, d) \in P$, and $\omega(b, d) \geq 0$ is a non-negative weight function. This function shows the features in the form of persistence images of the connected components at the exact location they have appeared.

### 3.6. Patch Extraction

A patch is extracted from the image containing connected components by taking the centroid of the connected component and extracting a patch of *128 X 128* from the image of dimensions *240 X 240*. Due to the sparseness of the connected components compared to the pixels in the image and variations in intensities, there may be multiple connected components within or near the tumor region. Therefore, the centroid is selected based on a connected component to obtain an accurate patch that includes the tumor region. Complete flow of 2D patching using topological data analysis is shown in figure 4.

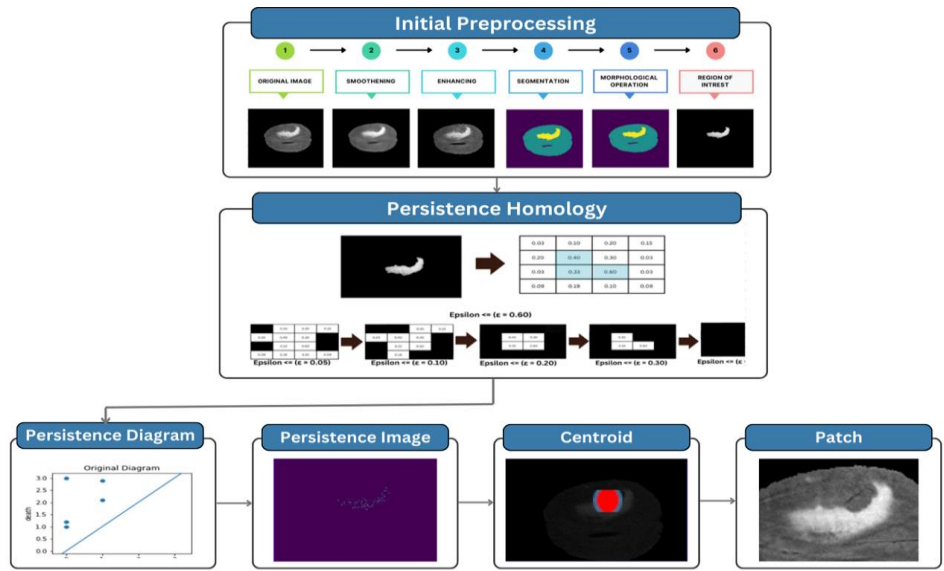

*Figure 4: Two Dimensional Patch Extraction*

### 3.7. Three Dimensional Patching using Connected Component Analysis

The brain MRI data in Brats 2020 is three-dimensional multimodal data. These three dimensions are coronal, sagittal and axial view of data, which are shown in figure 5. The data consist of multiple modalities out of these all modalities the preprocessing and centroid extraction is done on flair modality as the appearance of whole tumor is very clear on this modality. The process involved blurring and sharpening using customized filters and multilevel thresholding to extract the entire tumor region. This operation was performed on complete 3D volumes, considering the voxel information. After obtaining the 3D segmented ROI masks, connected component analysis was conducted, as illustrated in the figure 5. All the connected components were sorted based on the number of pixels they encompassed. To minimize noise from the extracted mask, we disregarded smaller connected components of fewer than 20 voxels, as determined by connected component labeling.

For evaluating the extracted mask, a comparison was made with the whole tumor label or the original segmentation mask using the Dice score. From the extracted mask, the centroid was identified, and a patch measuring 128x128x128 was extracted around that centroid.

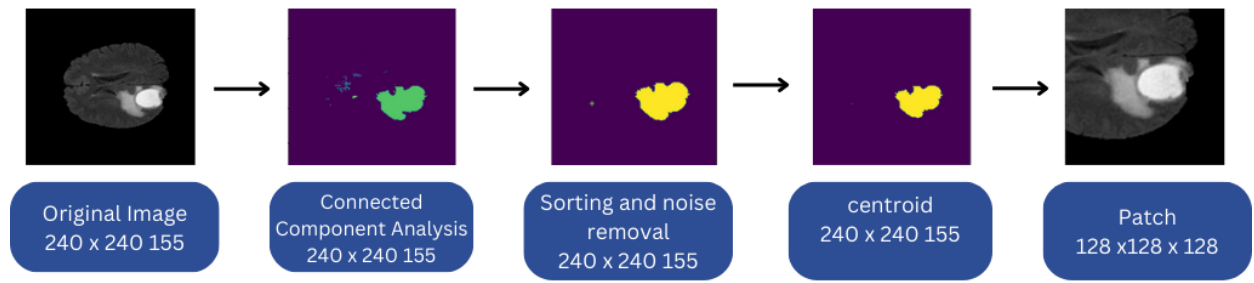

*Figure 5: Three Dimensional Patching*

### 3.8. Segmentation using U-NET

Experiments were carried out using the U-Net segmentation model to evaluate the effectiveness of this patching method in comparison to the other patching method as shown in figure 7. The fully connected neural network consists of an encoder and decoder network that minimize and maximize the dimensions, respectively for semantic segmentation. It consists of a skip connection and bridge, which helps the decoder layer learn better semantic features and connects the encoder and decoder layers.

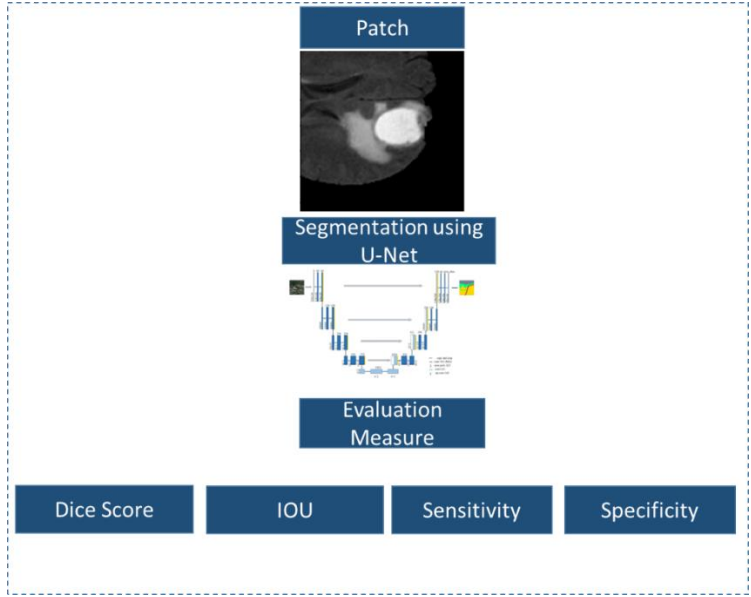

*Figure 6: Evaluation of patches by using U-NET*

## 4. Experiments and Results

In the experimentation phase, we aimed to validate the efficacy of our proposed patching technique and substantiate the contributions highlighted earlier. The experiments were meticulously designed to encompass a range of evaluations. Firstly, we conducted segmentation assessments, utilizing datasets generated through various patching techniques: fixed-size patching, random patching, fixed seed-based patching, random biased patching, and overlapping patching. This enabled us to gauge the accuracy of the segmentation process and the impact of each patching strategy on the overall results. Additionally, we explored the spatial relationship between the tumor center and the center of patches, employing the same set of patching techniques. This analysis provided valuable insights into how the different methods affect the alignment between patches and tumor regions. Furthermore, we investigated the issue of class imbalance and its mitigation using the diverse patching techniques, including our proposed approach. By comparing the results of these experiments, we were able to ascertain the effectiveness of each technique in addressing class imbalance. In conclusion, our experimentation and analysis shed light on the viability and advantages of our proposed patching technique, offering a comprehensive understanding of its impact on segmentation accuracy, spatial alignment, and class imbalance mitigation.

### 4.1. Dataset

The suggested technique underwent training using the BraTS 2020 training dataset (Bakas et al., 2017; Bakas et al., 2018). The training dataset consists of 369 scans obtained from patients, where each scan incorporates four distinct modalities: T1 weighted (T1), T2 weighted (T2), T1-Gadolinium (T1-GD), and fluid attenuated recovery (Flair). These scans are accompanied by segmentation masks, all of which adhere to the dimensions of (240, 240, 155). Conversely, the validation dataset encompasses 125 MRI scans, encompassing cases of both glioblastoma (GBM/HGG) and lower-grade glioma (LGG).For the purpose of evaluating segmentation performance, the analysis considers several specific sub-regions within the brain scans. Firstly, the "enhancing tumor" (ET) area is characterized by hyper-intensity in T1-Gadolinium (T1-GD) relative to T1, as well as when contrasted with "healthy" white matter in T1-GD. Secondly, the "tumor core" (TC) encompasses the majority of the tumor volume. This includes not only the enhancing tumor (ET) but also necrotic (fluid-filled) regions and non-enhancing (solid) components. Lastly, the "whole tumor" (WT) delineates the complete pathological extent within the brain scans. This involves both the tumor core (TC) and the surrounding peritumoral edema (ED), often depicted by hyper-intense signals in FLAIR images.The segmentation provided for this dataset employs specific values: 1 designates necrotic

(NCR) and non-enhancing tumor (NET) regions, 2 indicates peritumoral edema (ED), 4 represents the enhancing tumor (ET), and 0 is utilized to characterize all remaining areas within the scans. (Bakas et al., 2017; Bakas et al., 2018). An example showing the various regions of interest with the actual segmentation region can be seen in Figure 7. The details of samples are mentioned in the table 1.

*Table 1:Dataset Details*

| Patching technique | | No. of samples | Total 3D data comprising of all modalities |
|---|---|---|---|
| Single patch (Random Patch, Random Seed based Patch, CCA-based Patch) | Training | 276 | 1104 3D volumes |
| | Validation | 93 | 372 3D Volumes |
| | Testing | 125 | 500 3D volumes |
| Multiple patches (Fixed Size Patches, Overlapping Patches) | Testing | 276 x 4 = 1104 | 4416 3D volumes |
| | validation | 93 x 4 = 372 | 1488 3D volumes |
| | testing | 125 x 4 = 500 | 2000 3D volume |

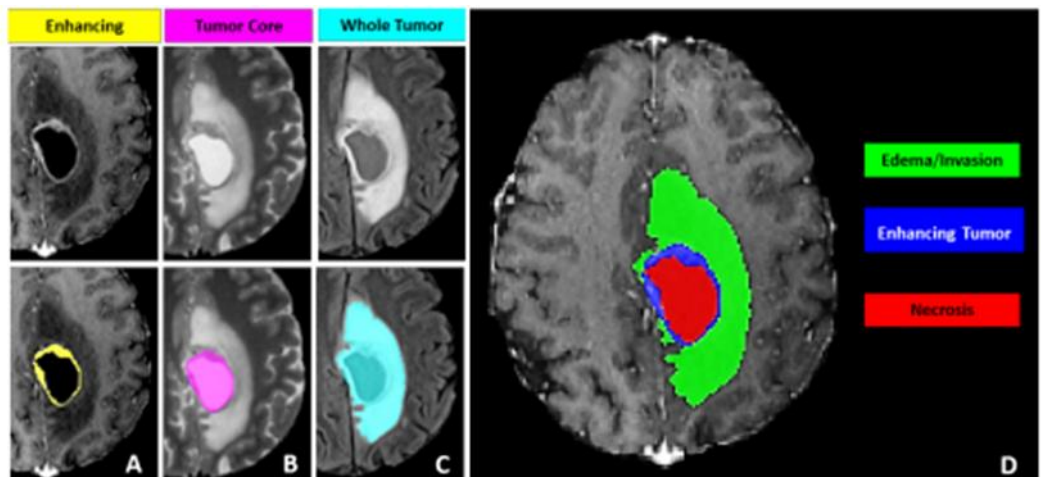

*Figure 7: Annotations on Brats 2020 on different modalities (A) Enhancing Tumor on T1ce (B) Tumor Core on Flair (c) Whole Tumor on T2 (D) Tumor Sub-regions wise annotations*

### 4.2. Experimental Setup

The proposed approach was implemented using the KERAS and TensorFlow with 16 GB NVIDIA P-100 GPU. Through experimentation, it is discovered that the most accurate segmentation is achieved by using a stack of volumes from different modalities. The input MR volumes are divided into 128 x 128 x 128 x 4 patches. The training of the model involves utilizing the ADAM optimizer (Kingma & Ba, 2014) with specific parameters (learning rate = 0.0001, epsilon = 1e−6). A batch size of 2 is employed, and the loss function is a combination of dice loss and focal loss, as described in equation 7 and 8 respectively. To prevent overfitting during the training process, dropout is implemented, with dropout rates set at 0.2 and 0.3. The model is constructed with a total of 5,646,260 parameters, which are utilized throughout the training procedure.

## 4.3. Evaluation Metrics

The following loss functions shown in euqation 7 and 8 are used of calculating the loss for acurate segmentation.

$$dice\ loss = 1 - \frac{2*\sum(p_p * p_t) + \epsilon}{\sum p_p^2 + \sum p_t^2 + \epsilon} \qquad (7)$$

Where $p_p$ is the predicted probablity map and $p_t$ is the true prbablity map and $\sum$ denotes the sum over all classes $\epsilon = epsilon$ is a small constant added to the denominator to avoid division by zero.

On the other hand, The Focal loss, as introduced in (Lin et al., 2017), was crafted with the aim of tackling the issue arising from imbalanced class distribution between foreground and background classes in the training process. It employs a parameterized function to reduce the impact of straightforward examples while amplifying the significance of challenging ones. Equation 8 provides the specific definition for the Focal loss function.

$$focal\ loss = -\alpha_t\ (1 - p_t)^\gamma\ \log(p_t) \qquad (8)$$

Here, $p_t$ represents the network probability of the positive class, α is the balancing parameter that controls the weight assigned to each class, and γ is the focusing parameter that adjusts the rate at which the weight assigned to easy examples is decreased. The modulating factor $(1 - p_t)^\gamma$ adjust the rate at which easy examples are down weighted.

The evaluation of proposed patch based unet model is done by using dice score as in equation 9, senitivity and spcifiity in equation 10 and 11 resectively, given by brats 2019, are defined as follows:

$$Dice = \frac{2*\sum(p_p * p_t) + \epsilon}{\sum p_p^2 + \sum p_t^2 + \epsilon} \qquad (9)$$

$$senitivity = \frac{TP}{TP + FN} \qquad (10)$$

$$specificity = \frac{TN}{TN + FN} \qquad (11)$$

where True Positive (TP), False Positive (FP), True Negative (TN) and False Negative (FN) denote the number of false negative, true negative, true positive, and false positive voxels.

## 4.4. Evaluation and Comparison with State-of-the-art-Techniques

We evaluated the patching technique through a comprehensive analysis of the center coordinates of patches in relation to the tumor centroid. This investigation aimed to demonstrate the effectiveness of the patching approach in positioning the tumor precisely at the center of each patch, thereby enhancing segmentation accuracy.

Additionally, we quantified the voxel count for each patching technique. This assessment enabled us to gauge the extent to which the proposed patching technique contributes to the reduction of class imbalance. For a visual representation of these findings, refer to Figure 8. The combined results of these evaluations provide valuable insights into the utility and impact of our proposed patching technique on improving segmentation precision and addressing class imbalance.

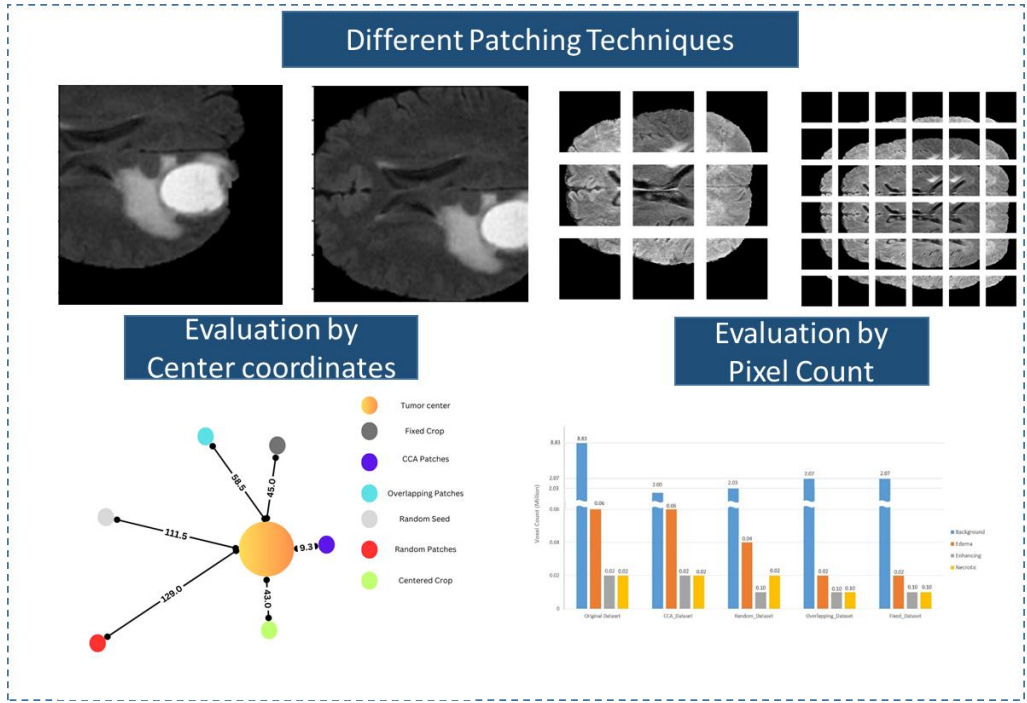

*Figure 8: Evaluation of different Patching techniques*

The Connected Component Analysis (CCA)-based 3D patching technique exhibited good performance in the segmentation task. This success stems from the positioning of the tumor at the center of the patch, enabling the model to effectively capture tumor-specific features. The technique also preserves tumor boundaries, resulting in segmentation outcomes that surpass those of other contemporary methods.

In contrast, random patching often leads to the omission of tumor regions, resulting in suboptimal tumor segmentation results. Notably, fixed patches and overlapping patches incurred significant training times, yielding outcomes that lacked effectiveness. This prolonged training time was primarily attributed to a fourfold increase in dataset size.

Furthermore, we assessed the effectiveness of other patching techniques by evaluating the distance of coordinates from the tumor center. Refer to Figure 9 for a visual representation of the method of these evaluations. These comprehensive findings collectively emphasize the superiority of the CCA-based 3D patching technique in achieving robust and accurate tumor segmentation results.

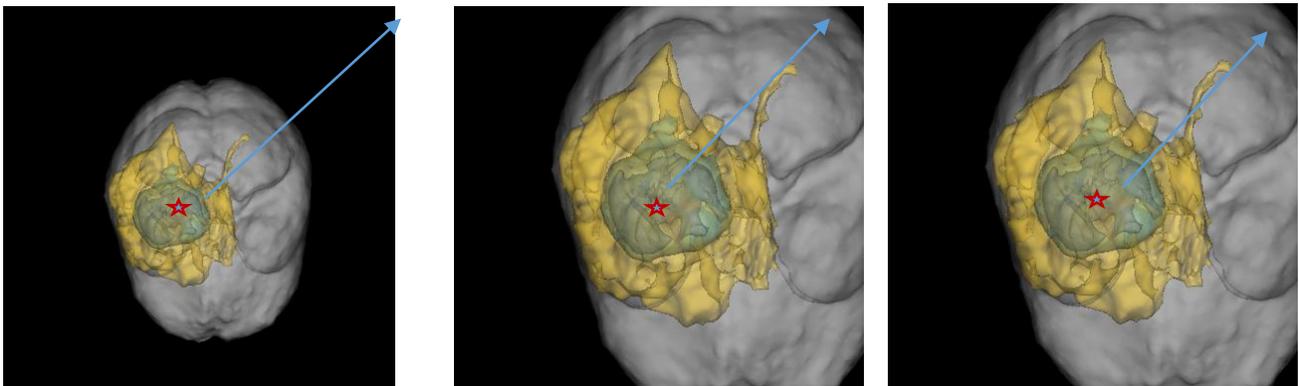

*Figure 9: (A) Original Mri(240 x 240 x 155) Distance from center of tumor to corner  (B) Distance of center coordinate of CCA patches (C) Distance of center coordinates of random patches*

The findings presented in Figure 10 provide a clear depiction of the analysis. The CCA-based patching technique demonstrates a notable trait of generating patches in close proximity to the tumor region. In contrast, the patches produced via random patching exhibit greater distance from the tumor center, leading to the omission of tumor segments and boundaries.

Similarly, the overlapping patches and centered crop patches also tend to be distant from the tumor center. This discrepancy arises because, in the majority of cases, the tumor does not manifest at the center of the original MRI; instead, it often appears on either side of the brain's lobes (Wang & Chung, 2022). These observations underscore the precision achieved by the CCA-based patching technique in accurately capturing tumor-related features and ensuring effective segmentation.

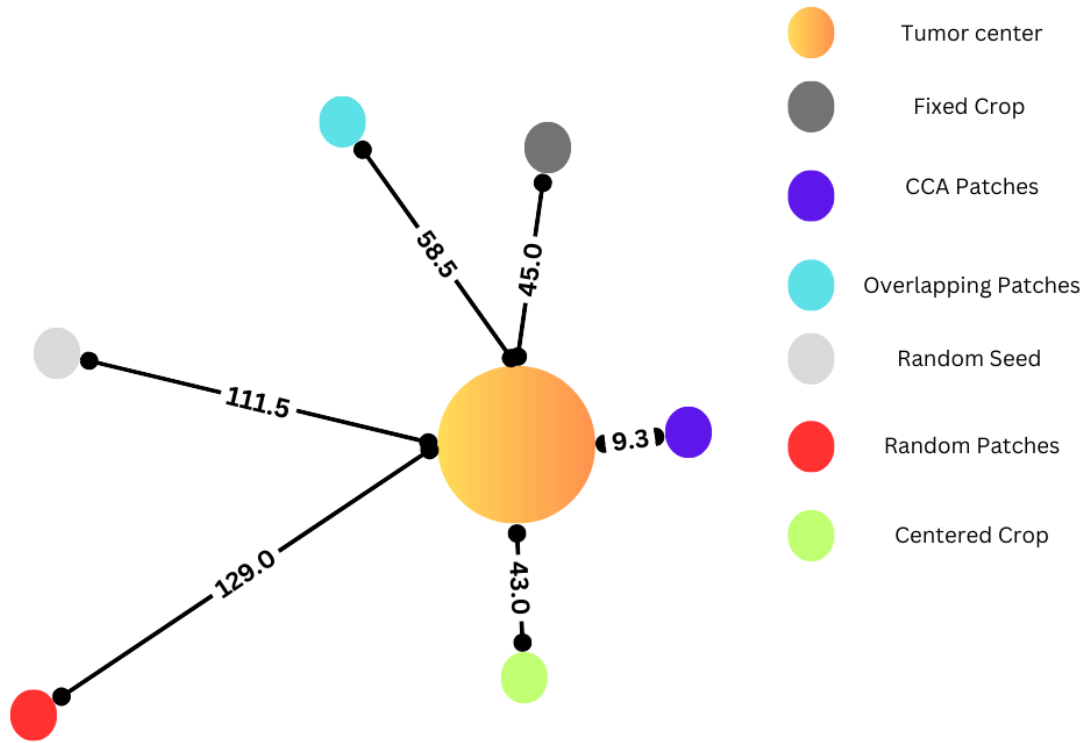

*Figure 10: Evaluation by measuring Distance from tumor center and center coordinates of patches*

We conducted an additional evaluation of the patches based on the percentage of tumorous voxels within the image. This approach was adopted due to the observation that tumor shape is not captured completely in other patching techniques, resulting in the exclusion of tumorous voxels.

In contrast, our patching technique centers the tumor within the patch, thereby preserving the tumor's shape while disregarding numerous background pixels. This strategy is particularly crucial due to the substantial class imbalance(Wang & Chung, 2022) present in the BraTS dataset. Our patching technique showcased a significant improvement in tumor representation, enhancing the tumor's visibility by fourfold compared to other patching techniques.For a visual representation of the original dataset's class imbalance, please refer to Figure 11.

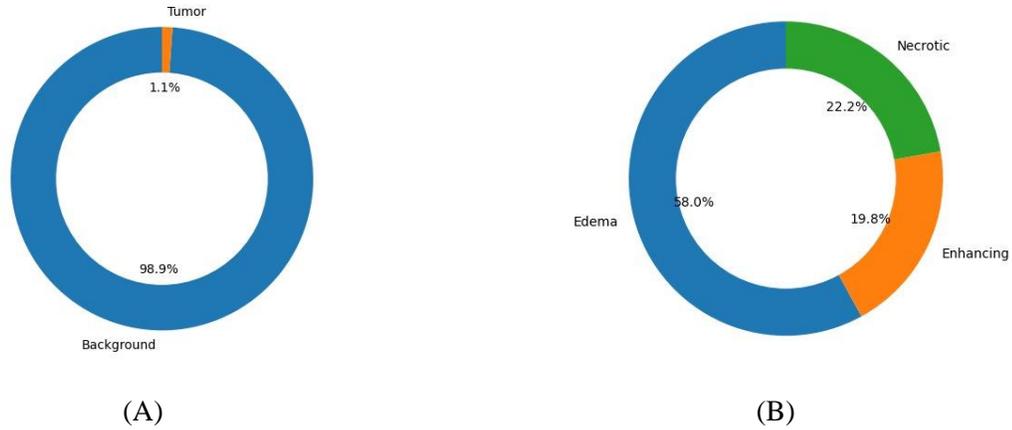

*Figure 11: Class Imbalance in Original Dataset (A) Imbalance in Tumor and Background pixels (B) Imbalance in tumor Sub-regions*

The reduction in class imbalance by using different patching techniques and the proposed technique has been shown in this figure 12. These outcomes affirm the efficacy of our patching technique in addressing class imbalance and enhancing the accurate representation of tumors within the dataset.

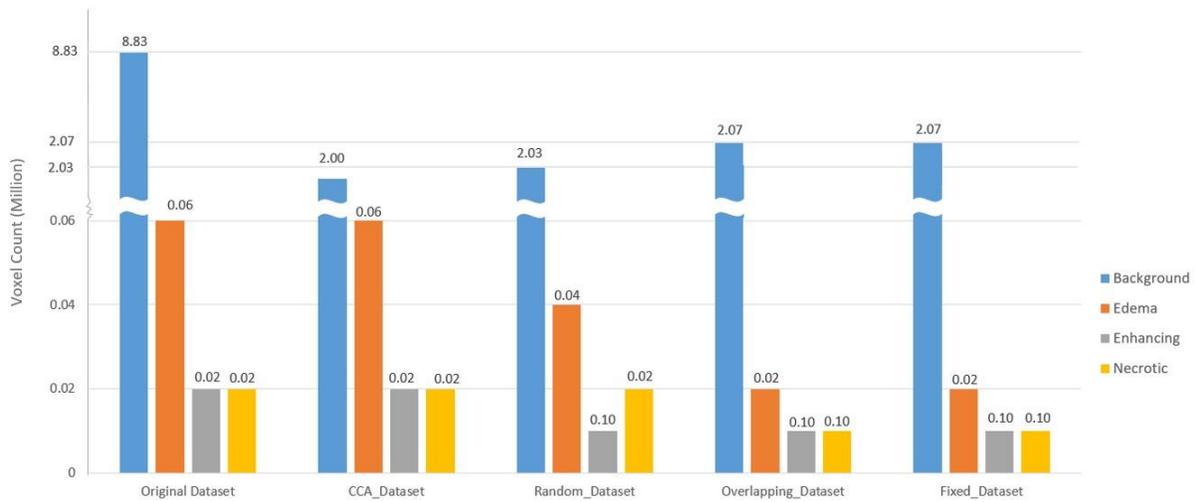

*Figure 12: Evaluation by Voxel count in State-of-the-art patching techniques*

.
## 4.5. Evaluation by Segmentation Results

The following results as shown in table 2 have been carried out on BraTs 2020 dataset of 3D volumes, various state of the art techniques have been used in comparison to our proposed methadology on segmentation task. The model and parameters used across each comparison were same with the same U-Net Model, optimizer and loss function and number of epochs.

*Table 2: Comparison with State-of-the-art patching techniques*

| Patching Technique | Dice | Edema | necrosis | WT | ET | TC | Time (h= Hours) |
|---|---|---|---|---|---|---|---|
| Centered Crop | 0.52 | 0.55 | 0.32 | 0.62 | 0.28 | 0.30 | 9.06 h |
| Fixed 26[th] epoch | 0.59 | 0.53 | 0.37 | 0.64 | 0.45 | 0.41 | 31.7 h |
|  | 0.54 | 0.36 | 0.22 | 0.52 | 0.21 | 0.21 |  |
| Random patching | 0.64 | 0.63 | 0.46 | 0.69 | 0.55 | 0.50 | 8.9 h |
| Patching with seed 42 | 0.63 | 0.60 | 0.42 | 0.67 | 0.54 | 0.48 | 8.86 h |
| Overlapping with a stride of 64 | 0.72 | 0.64 | 0.48 | 0.70 | 0.59 | 0.53 | 33.9 h |
| **CCA Based** | **0.74** | **0.73** | **0.61** | **0.78** | **0.71** | **0.76** | **8.7 h** |

The results obtained from the segmentation using our proposed patching technique demonstrate its superiority over other state-of-the-art patching methods. The model achieved impressive performance without any compromise in accuracy. Fixed patches, although increasing the dataset size and training time, did not match the results obtained by our proposed technique. Our approach successfully preserved tumor boundaries and maintained tumor voxels in the centered crop, leading to significantly improved results. Moreover, random patching showed some improvement due to fewer missing tumor voxels, but it could not outperform our method. Similarly, patching with a seed exhibited good results, although it still fell short of the performance achieved by our proposed approach. In contrast, overlapping patches had a negative impact on tumorous voxels in specific regions, affecting their segmentation results. Overall, our proposed model demonstrated exceptional performance, surpassing other techniques, while providing accurate and efficient tumor segmentation results

## 5. Conclusion

In conclusion, our research presents a novel patching technique that addresses two critical challenges in tumor segmentation. Firstly, it efficiently forms tumor-centered patches, effectively reducing class imbalance in the dataset. Secondly, it enables the use of the U-net architecture with significantly fewer parameters, optimizing performance without compromising accuracy.

The proposed architecture achieved a remarkable Dice score of 74.0% using simple lightweigh U-Net for the entire tumor segmentation, highlighting its effectiveness in correctly identifying tumor boundaries. By preserving boundaries through the placement of tumors at the center of the patches, our technique significantly improves the accuracy of tumor identification.

An essential advantage of our patching approach lies in its ability to counter the problem of class imbalance. Unlike traditional patching methods that lead to over-representation of the background class during training, our technique minimizes background pixels while maintaining the tumor shape intact. This ensures that the voxels in the tumor remain consistent with those in the original dataset without patching.In contrast to alternative patching techniques, which often result in reduced tumorous and background voxels, our proposed approach achieves a balanced and accurate segmentation. The comparative analysis depicted in the figure demonstrates the superiority of our method.

In conclusion, our novel patching technique provides a robust solution to class imbalance and segmentation challenges in medical imaging. By generating tumor-centered patches and preserving tumor boundaries, we significantly improve the performance of the U-net architecture, resulting in superior tumor segmentation accuracy. Our findings have significant implications for enhancing tumor detection and segmentation in medical imaging applications, ultimately contributing to improved diagnostic accuracy and patient care.

**Acknoledgement**
This work has been supported by Higher Education Commission under under Grand Challenge Fund (GCF-HEDP) with research project number GCF-912 and was carried out at Artificial Intelligence Diagnostics (AID) Lab at FAST-NUCES, Islamabad Campus in collaboration with Aga Khan University, Karachi.